\documentstyle[version2,aps]{revtex}
\begin{document}
\draft
\twocolumn
\widetext
\begin{title}
Heavy Quasiparticles in the Anderson Lattice Model
\end{title}
\author{K. Tsutsui$^1$, Y. Ohta$^2$, R. Eder$^3$, S. Maekawa$^1$,
E. Dagotto$^4$, and J. Riera$^5$}
\begin{instit}
$^1$Department of Applied Physics, Nagoya University,
Nagoya 464-01, Japan
\\
$^2$Department of Physics, Chiba University, Chiba 263, Japan
\\
$^3$Department of Applied and Solid State Physics,
University of Groningen, The Netherlands
\\
$^4$Department of Physics, National High Magnetic Field Laboratory
and Martech,\\
Florida State University, Tallahassee, FL 32306, USA
\\
$^5$Departamento de Fisica, Fac. de Cs. Exactas e Ingeniera,
2000-Rosario, Argentina
\end{instit}
\begin{abstract}
An exact-diagonalization technique on small clusters is used
to study the dynamics of the one-dimensional symmetric Anderson
lattice model.  Our calculated excitation spectra reproduce key
features expected for an infinite Kondo lattice such as
nearly localized low-energy spin excitations and extended regions
of `heavy-quasiparticle' bands.  We show that, in contrast to
the hybridization picture, low-energy spin excitations of
the nearly-localized $f$-electron system play a key role in the
formation of an almost dispersionless low-energy band of heavy
quasiparticles.
\end{abstract}
\narrowtext
\topskip9.3cm

The origin of the anomalous behavior of $f$-electron
compounds is an unresolved issue in the theory of
strongly-correlated electron systems.  Thereby the way in
which a periodic array of magnetic ions interacting with a sea
of conduction electrons can give rise to either the extreme
low-energy scale in the Landau-type quasiparticle bands of
heavy-Fermion compounds\cite{Fulde,Coleman} or to gaps of
apparent many-body origin in the excitation spectra of Kondo
insulators\cite{Cooley,Park} is not yet understood.
On a phenomenological level, heavy-Fermion compounds have been
described with considerable success by the `renormalized band
theory'\cite{Fulde} where the effect of electron correlations
is described by the renormalization of on-site energy and
hybridization strength of the magnetic ions.

In this Letter we study the Anderson lattice model (ALM),
the simplest model relevant for $f$-electron compounds,
by Lanczos diagonalization of small clusters\cite{Jullien}, and
show that such a renormalized band picture on one hand may
provide a reasonable phenomenological description of the
dispersion relations, but on the other hand is not really
adequate on a microscopic level: contrary to the one-particle
picture, the heavy quasiparticles may be viewed as loosely-bound
states of conduction electrons and spin-wave--like excitations
of the nearly-localized $f$-electron system.  The emerging
picture is thus more reminiscent of the spin polaron discussed
recently on the basis of a semiclassical treatment of the Kondo
lattice\cite{Tsvelik}.
We would like to stress that due to the small size of the clusters
our calculations can neither reproduce the exponentially small energy
scales present in the Anderson impurity problem
nor do they allow for the derivation of an accurate phase digram.
Nevertheless one may expect that the relative magnitudes of
energy scales and hence the nature of the low lying states
are reproduced correctly by our calculations; in that sense
our results for dynamical quantities are complementary to
the scaling theories in Ref. \cite{Varma}
or the renormalization group calculations for the ground states
of larger systems in Ref. \cite{Yu}.

We consider a tight-binding version of the one-dimensional ALM
defined by the Hamiltonian
                          \begin{eqnarray}
                       H=&-&t\sum_{<ij>\sigma}
             (c_{i\sigma}^\dagger c_{j\sigma}+{\rm H.c.})
     -V\sum_{i\sigma}(c_{i\sigma}^\dagger f_{i\sigma}+{\rm H.c.})
                           \nonumber \\
                &+&U\sum_i (n_{i\uparrow}^f-{1\over 2})
                   (n_{i\downarrow}^f-{1\over 2}),
                            \label{ham}
                           \end{eqnarray}
where $c_{i\sigma}$ ($f_{i\sigma}$) is the annihilation operator
for an electron of spin $\sigma$ at site $i$ in the $c$- ($f$-)
orbitals and $n_{i\sigma}^f$$=$$f_{i\sigma}^\dagger f_{i\sigma}$.
Model parameters are hopping strength $t$ between
nearest-neighbor $c$ orbitals, mixing $V$ between $c$ and $f$
orbitals, on-site repulsion $U$ at the $f$ orbitals.
The on-site energy of the $f$ orbitals is taken to be $-U/2$,
i.e., we consider the `symmetric' case.  We focus on electron
densities close to `half filling', i.e., $2N_s$ electrons in
$N_s$ unit-cells (a unit cell contains one $c$ and one $f$
orbital).  Restrictions on memory space and computer time
necessitate to choose $N_s$$\leq$$6$.  To get additional
information, we employ twisted boundary
conditions\cite{Riera,Poilblanc} (BC).  We require
$c_{N_s+1\sigma}$$=$$e^{i\varphi}c_{1\sigma}$ and
$f_{N_s+1\sigma}$$=$$e^{i\varphi}f_{1\sigma}$ by introducing
an arbitrary phase $\varphi$; the allowed momenta are then
$k$$=$$(2\pi n$$+$$\varphi)/N_s$ with $n$$=$$0,\dots,N_s$$-$$1$.
It has been pointed out\cite{Tsune,Nishino} that
in the half-filled case finite size effects
can be minimized by choosing $\varphi$$=$$(N_s/2)\cdot \pi$.
As will be seen below, combination of spectra obtained with
different values of $\varphi$ gives remarkably smooth `band
structures'.  However, none of our conclusions to be presented
below depends on such an assignment of bands.

\topskip0cm

We first consider the single-particle spectral function
$A_\gamma(k,\omega)$ defined as a sum
$A_\gamma(k,\omega)=A_\gamma^-(k,-\omega)
+A_\gamma^+(k,\omega)$ of the photoemission (PES) spectrum
                         \begin{eqnarray}
             A_\gamma^-(k,\omega)=\frac{1}{\pi} \Im \;
          \langle \Psi^{N}_{\varphi}|\gamma_{k\sigma}^\dagger
           \frac{1}{\omega - (H-E^N_{\varphi})- i\epsilon}
                         \gamma_{k\sigma}
                     |\Psi^N_{\varphi}\rangle,
                           \end{eqnarray}
and the inverse photoemission (IPES) spectrum
                           \begin{eqnarray}
             A_\gamma^+(k,\omega)=\frac{1}{\pi} \Im \;
          \langle \Psi^{N}_{\varphi}|\gamma_{k\sigma}
          \frac{1}{\omega - (H-E^N_{\varphi})- i\epsilon}
                        \gamma_{k\sigma}^\dagger
                        |\Psi^N_{\varphi}\rangle,
                             \end{eqnarray}
where $E^N_{\varphi}$ ($|\Psi^N_{\varphi}\rangle$)
denote the ground state energy (wave function) with $N$ electrons
and twisted BC of phase $\varphi$.
The operator $\gamma_{k\sigma}$ refers to the Fourier transform
of the operator for either conduction electrons ($c_{i\sigma}$)
or the $f$ electrons ($f_{i\sigma}$).  Results for
$A_\gamma(k,\omega)$ obtained by the standard Lanczos procedure
are given in Figure~1.  One can identify the `upper- and
lower-Hubbard bands' for the $f$ electrons, separated by an
energy $\sim$$U$.  They are dispersionless and somewhat
broadened, with almost pure $f$ character.  In addition to this
typical strong-correlation feature, there are two bands which
are more reminiscent of noninteracting electrons: one can
identify the unrenormalized $c$-electron band of width $4t$,
apparently split into two bands by hybridization with
a `renormalized' $f$ level in the middle of the Hubbard gap.
This feature first results in a well-defined gap between
PES and IPES spectra and second in extended regions of `heavy'
bands with apparently pure $f$ character.  The dispersion of
the two `hybridization bands' as well as the change from almost
pure $c$ character to almost pure $f$ character around $\pi/2$
thereby are both roughly consistent with the picture of
noninteracting electrons.  The spectral weight of the parts
with $f$ character, however, is substantially smaller than
in the parts with $c$ character; comparison shows that this
asymmetry is the more pronounced the larger the ratio $U/V$.

The change of $A_\gamma(k,\omega)$
with hole doping is at first
sight completely consistent with the picture of noninteracting
electrons\cite{Maekawa}: the chemical potential seems to shift
into the `heavy' band, so that a kind of Fermi surface emerges,
and upper- and lower-Hubbard bands remain unaffected.  In addition
to this rigid-band--like behavior, however, there is also a
modification of the `light' parts of the band structure, far from
$E_{\rm F}$: $c$-type spectral weight is transferred from
PES to IPES near $\pi/2$, i.e., the Fermi momentum for a
half-filled band of unhybridized conduction electrons.  The
change of $A_c(k,\omega)$ thus
is reminiscent of {\it unhybridized} conduction electrons.
On a phenomenlogical level, this could be reproduced if one assumed
that the `renormalized'
$f$-level energy is pinned near the chemical potential of
$N$$-$$N_s$ unhybridized conduction electrons, i.e., the Fermi
energy of a `frozen-core' band structure.

We now want to clarify the nature of the heavy-band states.
Important information can be obtained from the momentum
distribution function
$n_{\gamma\sigma}(k)$$=$$\langle\gamma_{k\sigma}^\dagger
\gamma_{k\sigma}\rangle$; more precisely, we study the change
of $n_{\gamma\sigma}(k)$ upon removing one electron.  In a six
unit-cell system, we evaluate the difference
$\Delta n_{\gamma\sigma}(k)$ between the $n_{\gamma\sigma}(k)$
in the lowest state with 5 down-spin and 6 up-spin electrons at
the total momentum $k_{\rm tot}$ and the $n_{\gamma\sigma}(k)$
of the ground state at half-filling.  We choose $k_{\rm tot}$
such that the single-hole state belongs to the `heavy' part of
the band.  In the hybridization model, the creation operators
in the lower hybridization band would read
$a_{k\sigma}^\dagger=
u_k c_{k\sigma}^\dagger+v_k f_{k\sigma}^\dagger$,
so that
$\Delta n_{c\uparrow}(k)=0$,
$\Delta n_{c\downarrow}(k) =-|u_k|^2 \delta_{k,-k_{\rm tot}}$,
$\Delta n_{f\uparrow}(k)=0$, and
$\Delta n_{f\downarrow}(k)=-|v_k|^2 \delta_{k,-k_{\rm tot}}$.
Since one may expect $u_k$$\approx$$0$ and $v_k$$\approx$$1$
in the heavy band, the electron is removed only from the $f$
species with spin down and at $k$$=$$-$$k_{\rm tot}$.
The calculated results for $\Delta n_{\gamma\sigma}(k)$ are
shown in Figure~2, where we note the following features, almost
all of which are in contrast to these predictions:
\\
(i) Independently of the actual momentum $k_{\rm tot}$ of the
single-hole state, $c$ electrons
of both spin directions are removed at the two $k_{\rm F}^c$,
\\
(ii) The resulting loss of up-spin electrons is compensated
by an almost $k$-independent spin polarization of the $f$
electrons,
\\
(iii) As the only agreement with the hybridization model there
is an extra `dip' in $n_{f\downarrow}(k)$ for
$k$$=$$-$$k_{\rm tot}$, which however diminishes rapidly in magnitude
for decreasing $V/U$.
\\
These results establish first of all that the `heavy quasiparticle'
is predominantly a `missing $c$ electron' with only small
admixture of $f$ character (for large $U/V$).
By contrast, the pure $f$ character of the lower Hubbard band
suggests that it is in this band where an $f$ electron is missing.
We thus have an energy separation of $c$-like and $f$-like
degrees of freedom (of order $U/2$), in contrast to the
hybridization scenario.  However, there must be some mechanism which
renders the missing $c$ electron `invisible' in
$A_c(k,\omega)$ when $k$ is in the `heavy' band.

As for this latter issue, we note that the spin polarization
of the $f$ electrons suggests the presence
of a spin excitation.  We therefore consider the
spin-excitation spectrum
                         \begin{eqnarray}
              S_\alpha(q,\omega)= \frac{1}{\pi} \Im \;
          \langle \Psi^{N}_{\varphi}|S_{\alpha q}^-
         \frac{1}{\omega - (H-E^N_{\varphi})- i\epsilon}
                          S_{\alpha q}^+
                      |\Psi^N_{\varphi}\rangle,
                          \end{eqnarray}
where $S_{\alpha q}^+$ is the Fourier transform of either the
total-spin raising operator $c_{i\uparrow}^\dagger c_{i\downarrow}
+f_{i\uparrow}^\dagger f_{i\downarrow}$ ($\alpha$$=$${\rm tot}$)
or the $f$-electron spin raising operator
$f_{i\uparrow}^\dagger f_{i\downarrow}$ ($\alpha$$=$$f$).
The calculated spectra (see Figure~3) show strong low-energy
peaks with negligible dispersion, which probably are
the (almost) local singlet-triplet excitations expected
for a Kondo lattice.  In $S_f(q,\omega)$, these low-energy peaks
are enhanced whereas the smaller peaks at higher energies
are suppressed: obviously the spin-flip of an $f$ electron in the
ground state to excellent approximation produces another
eigenstate.  There is a pronounced $k$ dependence of the
peak intensity, similar to spin waves in an antiferromagnet.
One may assume that this reflects the antiferromagnetic spin
correlations due to the RKKY-type interaction.  We also study the
change of the momentum distribution of the half-filled system due
to a spin excitation; more precisely we consider the difference
between the momentum distribution for the lowest state with
$S_z$$=$$1$ and momentum $\pi$ (i.e., the final state for the
low-energy peak in $S_\alpha(q,\omega)$) and that for
the ground state.  This difference is shown in the inset of Figure~3.
Whereas the $c$ electrons remain virtually unaffected
by the spin excitation, there is an almost $k$-independent
polarization of the $f$ electrons, as one would expect it
for a quantum spin system without charge degrees of freedom.
Again we find a remarkable degree of separation of the $c$- and
$f$-electron `subsystems', which may also provide a natural
explanation for the strongly different spin and charge excitations
found in previous studies
\cite{Tsune,Nishino,Vekic,Yu,Ueda,Santini,Callaway,Bucher,Severing}
of Kondo insulators.

Let us now combine the above results to obtain a simple picture
of the heavy states. Since they represent the parts of the PES/IPES
 spectrum with the lowest excitation energy,
let us consider the limit $V$$\rightarrow$$0$ and ask `How can we
remove or add an electron so as to lower the energy most efficiently?'.
In the half-filled ground state, there is on the average one
$f$-electron per unit cell, with only a small admixture of the
empty or doubly occupied $f$ site.  Removing or adding an $f$-electron
will on the average raise the energy by $U/2$, and thus is
unfavorable. Accordingly, the state $f_{k\sigma}|\Psi^N_{\varphi}\rangle$,
which would be the most natural ansatz within the hybridization
picture, has only small overlap with the true `heavy state',
particularly in the strong correlation case (i.e. small $V/U$).
One measure for the weight of this state in the ground state
would be the `depth' of the dip in $n_{f\sigma}(k)$.
On the other hand, a $c$-electron can be removed or added
with practically no cost in energy if that is done near $k^c_{\rm F}$.
Next, the $f$-electron spin excitations with their
small excitation energies offer a way to dispose of `excess
momentum' with almost no cost in energy.
This suggests to remove or add the $c$-electron always at
$k^c_{\rm F}$, and transfer the excess
momentum to an $f$-spin excitation. This picture immediately explains
the reduction of $n_{f\sigma}(k^c_{\rm F})$, as well as
the spin polarization of the $f$-electron system due to
the accompanying $f$-spin excitation. We are thus led to the
following ansatz for a hole-like `heavy state':
                       \begin{eqnarray}
        |\Psi(k)\rangle = \biggl\{u_k f_{-k\downarrow} &+&
           \sum_{k^c_{\rm F}} v_{k^c_{\rm F}}
         \Bigl[c_{k^c_{\rm F}\downarrow} S_f^z(k+k^c_{\rm F})
                        \nonumber \\
      &-&c_{k^c_{\rm F}\uparrow} S_f^+(k+k^c_{\rm F})\Bigr]\biggr\}
                      |\Psi^N_{\varphi}\rangle.
                         \label{qp}
                       \end{eqnarray}
Here $S_f^z(q)$ is the $z$-spin operator for the $f$-electrons
with momentum transfer $q$, and $u_k$ and $v_k$ are (variational)
parameters.  The state (Eq.~(\ref{qp})) has momentum $k$,
$z$-spin $1/2$, and total spin $S$$=$$1/2$, i.e., the spins of
$f$-electron excitation and $c$-electron hole maximally compensate
each other.  This is reminiscent of the `quenching' of a
Kondo-impurity spin due to bound-state formation.
The variational parameters in (\ref{qp}) are determined from the
requirement that $|\Psi(k)\rangle$ has norm $1$ and maximum overlap
with the exact `heavy' state with momentum $k$.
Figure~4 shows the overlap
$|\langle \Psi (k) | \Psi^{N-1}_\varphi \rangle |^2$ for different
values of $V/t$ and $U/t$ at $k$$=$$5\pi/6$;
here $|\Psi^{N-1}_\varphi \rangle$ denotes the exact `heavy' state
and $|\Psi (k) \rangle$ is given by Eq.~(\ref{qp}).  For comparision,
the overlap of the state $f_{-k\downarrow}|\Psi^N_\varphi \rangle$
(normalized to unity) with
$|\Psi^{N-1}_\varphi \rangle$ is also shown
(in the hybridization picture, the latter quantity would be $1$).
While the `bare $f$-electron' is a good approximation
only in the small $U/V$ case,
the overlap of the state in Eq.~(\ref{qp}) is $>$$90$$\%$,
for all parameter values,
so that we find a good description of the heavy-band states
in the strong correlation region.

In summary, we have studied the single-particle spectral
function and dynamical spin-correlation function for finite
clusters of the Anderson lattice model at and near half filling.
Despite the necessarily rather coarse energy scales available
in the clusters, our results do reproduce key features expected
for infinite Kondo lattices, namely extended heavy bands and
almost dispersionless low-energy spin excitations.
On a phenomenological level, the low-energy
parts of the spectral function can be described reasonably well
by a renormalized band picture where an `effective $f$ level'
pinned to the `frozen-core' Fermi energy mixes with the conduction
band.  This picture, however, has not much significance beyond a purely
phenomenological level: there is a clear separation between the
low-energy `hybridization bands' which correspond to a missing
(or extra) $c$ electron and the two `Hubbard bands' which
correspond to a missing (or extra) $f$ electron.
The heavy quasiparticles rather have the character of loosely bound
states between a conduction electron at the Fermi momentum of
the unhybridized conduction-electron system and a spin-wave--like
excitation of the $f$-electron lattice which acts very much
like a pure quantum-spin system.

Since the `spin polaron bands' are formed by bound states, rather
than by true hybridization, it seems natural to assume that the
breaking of the bound states will completely remove the heavy
parts of the band structure and leave behind only the frozen-core
Fermi surface.  Since the heavy quasiparticles involve the
spin compensation of $f$ excitation and $c$ hole, it is moreover
clear that they can be broken by a magnetic field.  Then, the
breaking of the heavy polarons by a magnetic field and the
corresponding collapse of the Fermi surface to the frozen-core
volume then appears as a natural explanation for the so-called metamagnetic
transition associated with `itinerant-to-localized' nature of
$f$ electrons\cite{Aoki}.

We thank S.~Haas for useful conversations.  This work was supported
by Priority-Areas Grants from the Ministry
of Education, Science, and Culture of Japan, the New Energy and
Industrial Technology Development Organization (NEDO), and the
Office of Naval Research under Grant No. ONR-N00014-93-1-0495.
Support of R. E. by the European Community is acknowledged.
Computations were carried out in the Nagoya University Computation
Center, the Computer Center of Institute for Molecular Science,
Okazaki National Research Institutes, and the Cray C90 at Pittsburg
Supercomputer Center.


\figure{Single-particle excitation spectra $A_\gamma(k,\omega)$
for $N_s$$=$$6$ with different $\varphi$ for
$V/t$$=$$1$ and $U/t$$=$$6$.
The spectra at $k$$=$$0$, $\pi/3$,
$2\pi/3$, and $\pi$ correspond to $\varphi$$=$$0$ (i.e., periodic
BC).  The $f$ spectra are multiplied by $-$$1$
for better distinction, the
Lorentzian-broadening is $\epsilon$$=$$0.02t$.
The upper panel shows the spectra for the half-filled ground state,
the lower panel for the ground state
with 5 up and 5 down electrons (i.e., with two
holes).  The vertical dashed line shows the chemical potential.
\label{fig1}}

\figure{Difference $\Delta n_{\gamma\sigma}(k)$
between zero-hole and
one-hole ground states at $U/t$$=$$6$;
(a) $V/t$$=$$1$, $k_{\rm tot}$$=$$5\pi/6$,  $\varphi$$=$$\pi$
(i.e., antiperiodic BC),
(b) $V/t$$=$$1$, $k_{\rm tot}$$=$$\pi$, $\varphi$$=$$0$
(c) $V/t$$=$$0.5$, $k_{\rm tot}$$=$$5\pi/6$, $\varphi$$=$$\pi$.
As a reference $n_{\gamma\sigma}(k)$ for the zero-hole ground
state at $V/t$$=$$1$ is shown in (d).
\label{fig2}}

\figure{Spin-excitation spectra $S_\alpha(q,\omega)$ for
$V/t$$=$$1$ and $U/t$$=$$6$, $\varphi$$=$$0$.
The Lorentzian-broadening $\epsilon$$=$$0.02t$.
Inset shows the change of $n_{\gamma\sigma}(k)$
due to the spin excitation.
\label{fig3}}

\figure{Overlap of the state given by Eq.~(\ref{qp}) (white symbols)
and the state
$f_{-k\downarrow}|\Psi^N_{\varphi}\rangle/
\langle \Psi^N_{\varphi}|
f_{-k\downarrow}^\dagger
f_{-k\downarrow}|\Psi^N_{\varphi}\rangle^{1/2}$ (black symbols)
with the exact `heavy'
state $|\Psi^{N-1}_\varphi\rangle$ at $k$$=$$5\pi/6$ and
$\varphi$$=$$\pi$ as functions of $U/t$.
\label{fig4}}
\end{document}